\documentclass[9pt,twocolumn,twoside]{opticajnl}
\journal{opticajournal} 

\setboolean{shortarticle}{false}

\usepackage{bm, color}
\usepackage{multirow}
\usepackage{array, cases, amsmath, physics}
\usepackage{hyperref}

\newcommand{\spD}[1]{\fn{\tilde{\chi}_{_V}}{#1}}

\newcommand{\tens}[1]{\boldsymbol{#1}}

\newcommand{\uvect}[1]{\hat{\vect{#1}}}

\newcommand{\fn}[2]{\mathinner{#1\mathopen{\left(#2\right)}}}
\newcommand{\vect}[1]{{\bf #1}}
\newcommand{\E}[1]{\left\langle#1\right\rangle}

\newcommand{\F}[1]{\fn{F^\mathrm{(1D)}}{#1}}

\usepackage{soul}
\setstcolor{red}

\setul{0}{0.3ex}

\usepackage[colorinlistoftodos,shadow,textwidth=18mm]{todonotes}

\begin{document}

\title{Effective electromagnetic wave properties of disordered stealthy hyperuniform layered media beyond the quasistatic regime}

\author[1,2,3]{Jaeuk Kim}
\author[1,2,3,4,*]{Salvatore Torquato}
\affil[1]{Department of Chemistry, Princeton University, Princeton, New Jersey 08544, USA}
\affil[2]{Princeton Materials Institute, Princeton University, Princeton, New Jersey 08544, USA}
\affil[3]{Department of Physics, Princeton University, Princeton, New Jersey 08544, USA}
\affil[4]{Program in Applied and Computational Mathematics, Princeton University, Princeton, New Jersey 08544, USA}
\affil[*]{torquato@princeton.edu}

\begin{abstract}
    \emph{Disordered stealthy hyperuniform} dielectric composites exhibit novel electromagnetic wave transport properties in two and three dimensions. 
    Here, we carry out the first study of the electromagnetic properties of one-dimensional (1D) disordered stealthy hyperuniform layered media.   
    From an exact nonlocal theory, we derive an approximation formula for the effective dynamic dielectric constant tensor $\fn{\tens{\varepsilon}_e}{\vect{k}_q,\omega}$ of general 1D media that is valid well beyond the quasistatic regime and apply it to 1D stealthy hyperuniform systems. 
    We consider incident waves of transverse polarization, frequency $\omega$, and wavenumber $k_q$.
    Our formula for $\fn{\tens{\varepsilon}_e}{\vect{k}_q,\omega}$, which is given in terms of the \emph{spectral density}, leads to a closed-form relation for the transmittance $T$. 
    Our theoretical predictions are in {excellent} agreement with finite-difference time-domain (FDTD) simulations.  
    Stealthy hyperuniform layered media have perfect transparency intervals up to a finite wavenumber, implying no Anderson localization, but non-stealthy hyperuniform media are not perfectly transparent.  
    Our predictive theory provides a new path for the inverse design of the wave characteristics of disordered layered media, which are readily fabricated, by engineering their spectral densities.  
\end{abstract}
\maketitle

\section{Introduction} \label{sec:intro}

    \emph{Disordered hyperuniform} many-body systems \cite{To03a, Za09, To18a} are an emerging class of amorphous states of matter that are endowed with the novel wave and other transport properties with advantages over their periodic counterparts {\color{black}\cite{Fl09b, Man13b, Ma16, Le16, Zh16b, Xu17, Fr17, Kl18, torquato_extraordinary_2022, Ch18a, Zhang18b, To18a, Go19,  Ki20a, yu_Engineered_2021, romero-garciaWaveTransport1D2021, sgrignuoli_Subdiffusive_2022, granchi_NearField_2022, tavakoli_65_2022, cheron_Wavetransportstealth_2022,torquato_extraordinary_2022, klatt_Wavepropagationband_2022}.}
    Such hyperuniform two-phase composites are characterized by an anomalous suppression of volume-fraction fluctuations in the infinite-wavelength limit \cite{Za09, To18a}, i.e., the local volume-fraction variance $\fn{\sigma_V^2}{R}$ inside a spherical observation window of radius $R$ decays faster than $R^{-d}$ in $d$ dimensions in the large-$R$ limit, $\lim_{R\to\infty} R^d \fn{\sigma_V^2}{R} = 0.$
    Equivalently, its associated \emph{spectral density} $\spD{\vect{k}}$ vanishes as the wavenumber $\abs{\vect{k}}$ tends to zero, i.e., $\lim_{\abs{\vect{k}}\to0}\spD{\vect{k}} = 0.$
    One important class of such hyperuniform media are the disordered \emph{stealthy} varieties in which $\spD{\vect{k}}=0$ for $0<\abs{\vect{k}}<K$ \cite{Uc04b, Ba08, Zh15a, To15}, meaning that they completely suppress single scattering of incident radiation for these wave vectors \cite{Ba08,To18a}.  
    The degree of stealthiness $\chi$ is the ratio of the number of the constrained wave vectors in the reciprocal space to the total number of degrees of freedom.
    Recent studies showed that such exotic disordered media exhibit novel electromagnetic wave transport properties, including high transparency in the optically dense regime, maximized absorption, and complete photonic band-gap formation {\color{black}\cite{Fl09b, Man13b, Le16, Zh16b, Fr17, torquato_extraordinary_2022, Ch18a, Go19, Zhang18b, zhou_UltraBroadbandHyperuniformDisordered_2020, Kl20a, Ki20a, sheremet_absorption_2020-1, yu_Engineered_2021, granchi_NearField_2022, tavakoli_65_2022,  cheron_Wavetransportstealth_2022, torquato_extraordinary_2022, klatt_Wavepropagationband_2022},} in two and three spatial dimensions. 
    For example, previous work on light transparency of stealthy hyperuniform systems considered 2D point scatterers \cite{Le16} and 2D and 3D two-phase dielectric composites \cite{Fr17,torquato_nonlocal_2021}.
   Here, we undertake the first study of the electromagnetic properties of 1D disordered stealthy hyperuniform layered media.
    
    The problem of wave propagation in a two-phase (or multi-phase) layered medium has been extensively studied,  because of its simplicity and its ease of fabrication \cite{elser_Nonlocaleffectseffectivemedium_2007, bhattacharjee_Recenttrendsmultilayered_2017}, including surface plasmons \cite{maierPlasmonicsFundamentalsApplications2007, hohenester_StratifiedMedia_2020}, Anderson localization \cite{anderson_AbsenceDiffusionCertain_1958, mcgurn_Andersonlocalizationonedimensional_1993, Sh95, aegerter_Coherent_2009, izrailev_anomalous_2012,   wiersma_disordered_2013}, and deep-subwavelength disorder \cite{herzigsheinfux_Interplayevanescencedisorder_2016}, as well as practical applications \cite{salandrino_Farfieldsubdiffractionoptical_2006, xiong_Projectingdeepsubwavelengthpatterns_2008,  romero-garciaWaveTransport1D2021, oh_Controllocalizationoptical_2022}.
    An important wave characteristic is the effective dynamic dielectric constant tensor $\fn{\tens{\varepsilon}_e}{\vect{k}_q, \omega}$ for the incident radiation of frequency $\omega$ and wave vector $\vect{k}_q$.
    This complex-valued quantity determines the effective wavenumber $k_e$ and extinction mean free path $\ell_e$. 
    While there have been many theoretical/numerical treatments to estimate $\fn{\tens{\varepsilon}_e}{\vect{k}_q, \omega}$ of layered media \cite{yeh_optical_2005-1, rytov_Electromagneticpropertiesfinely_1956,
    sjoberg_ExactAsymptoticDispersion_2006, maurel_effective_2008, chebykin_Nonlocaleffectivemedium_2011, popov_Operatorapproacheffective_2016, merzlikin_HomogenizationMaxwellequations_2020, wen_nonlocal_2021}, previous approximations are applicable to disordered media in the \emph{quasistatic} or long-wavelength regime, i.e., $\abs{\vect{k}_q}\xi \ll 1$, where $\xi$ is a characteristic inhomogeneity length scale.


    Torquato and Kim \cite{torquato_nonlocal_2021}  recently derived the general nonlocal strong-contrast expansion for the effective dielectric constant tensor $\fn{\tens{\varepsilon}_e}{\vect{k}_q, \omega}$ that can be applied to two-phase media with various symmetries.  
    This expansion exactly treats multiple scattering to all orders beyond the quasistatic regime (i.e., $0\leq \abs{\vect{k}_q}\xi \lesssim 1$)  as a series involving functionals of the $n$-point correlation functions $\fn{S^{(i)}}{\vect{x}_1,\ldots,\vect{x}_n}$ for all $n$ (Section \ref{sec:theory}).
    Here, the quantity $\fn{S^{(i)}}{\vect{x}_1,\ldots,\vect{x}_n}$ gives the probability of finding $n$ points at positions $\vect{x}_1,\ldots,\vect{x}_n$ all in phase $i\hspace{2pt}(=1,2)$.
    Because of the fast-convergence property of this series (or the linear fractional form of this series), truncating it at the $n$-point level yields  \emph{multiple-scattering} approximations that still accurately capture multiple scattering to all orders for a wide class of microstructures, including statistically \emph{anisotropic} media. 
    The second-order truncations already provide accurate approximations for 2D and 3D statistically isotropic two-phase media \cite{torquato_nonlocal_2021}. 
    However, and importantly, analogous approximations for statistically anisotropic media have yet to be extracted and applied.
    
\begin{figure}[th]
    \centering
    \includegraphics[width=0.49\textwidth]{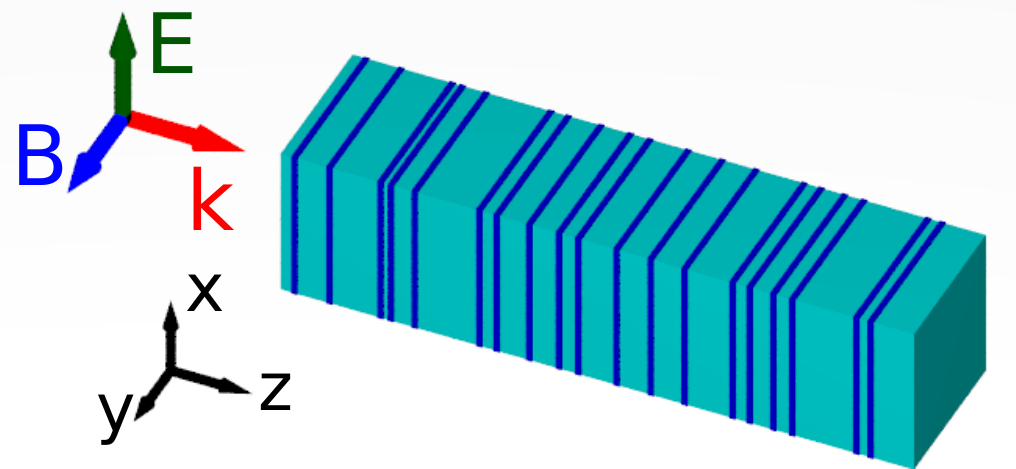}
    \caption{
        Schematic of three-dimensional disordered anisotropic stratified media consisting of infinite parallel slabs of phases 1 (cyan) and 2 (dark blue). 
        A plane electromagnetic wave of transverse polarization is normally incident into the medium, and its wave vector is shown as a red arrow.
        \label{fig:schematics}
    }
\end{figure}

    Here, we theoretically and numerically investigate the tensor $\fn{\tens{\varepsilon}_e}{\vect{k}_q, \omega}$ of 3D anisotropic layered media consisting of infinite parallel dielectric slabs of phases 1 and 2 whose thicknesses are derived from 1D disordered stealthy hyperuniform packings at various $\chi$ values.
    For simplicity, we focus on normally incident waves of transverse polarization, where $\vect{k}_q = k_q \uvect{z}$ (see Fig. \ref{fig:schematics}), and thus the wavenumber $k_q$ is the independent variable of the effective dielectric constant.
    From the exact strong-contrast expansion, we derive,  for the first time, formulas for $\fn{\tens{\varepsilon}_e}{k_q, \omega}$  for 3D anisotropic layered media that accurately accounts for multiple scattering in terms of the spectral density $\spD{\vect{k}}$,  enabling us to probe a wide range of wavenumbers. 
    The quantity $\spD{\vect{k}}$  is the Fourier transform of the autocovariance function $\fn{\chi_{_V}}{\vect{r}}\equiv \fn{S_2^{(i)}}{\vect{r}}-{\phi_i}^2$ \cite{To02a},  where $\vect{r}\equiv \vect{x}_2-\vect{x}_1$, and can be measured from scattering experiments \cite{De57}.  
    To our knowledge, this expression is the first closed-form formula of $\fn{\tens{\varepsilon}_e}{k_q, \omega}$ for general 1D media that applies well beyond the quasistatic regime. 
    
    We numerically verify that our derived formula can accurately capture multiple scattering effects due to correlated disorder beyond the quasistatic regime by using finite-difference time-domain (FDTD) simulations. 
    For this purpose, we numerically generate stealthy hyperuniform stratified dielectric two-phase media via a modified collective-coordinate procedure described in Refs. \cite{Ki20a,torquato_nonlocal_2021}  (Section \ref{sec:models}). 
    For dimensionless wavenumbers up to $k_1/\rho \lesssim 1.5$,  our predictions indeed show excellent agreement with the real and imaginary parts of the effective dielectric constant as well as transmittance found from FDTD simulations (Secs. 4-5 of \href{url}{Supplement 1}).
    Notably, our formula predicts that stealthy hyperuniform layered media are perfectly transparent (defined as $\Im[\fn{\varepsilon_e}{k_q,\omega}]=0$)  up to a finite wavenumber $K_T = K/(2\sqrt{\phi_1\varepsilon_1+\phi_2\varepsilon_2})$ (i.e., no Anderson localization) in the infinite-volume limit;  see \eqref{eq:trans-regime}.
    This result is especially remarkable because extended states in 1D  disordered systems are more difficult to achieve than in higher dimensions \cite{Sh95, aegerter_Coherent_2009, izrailev_anomalous_2012, wiersma_disordered_2013, yu_Engineered_2021}.
    
    We also show that a perfect transparency interval cannot exist in disordered 1D \emph{non-stealthy} media, {\it hyperuniform} or not, and thus Anderson localization can be present at all wavenumbers.  
    Our results, combined with the methods to generate media with a prescribed spectral density \cite{Uc04b, Ba08, Zh15a, Ch18a}, provide a new inverse-design approach \cite{To09a} to engineer and fabricate multilayered dielectric media with novel wave properties.

\section{Theory}
\label{sec:theory}

\subsection{Exact Strong-Contrast Expansion}    \label{sec:general}

Here, we briefly summarize the general nonlocal strong-contrast-expansion formalism of the effective dynamic dielectric constant tensor $\fn{\tens{\varepsilon}_e}{\vect{k}_q, \omega}$  for 3D two-phase media with arbitrary symmetries \cite{torquato_nonlocal_2021}. 
(The strong-property-fluctuation theory \cite{tsang_scattering_1981,mackay_thirdorder_2001} corresponds to a special case of our strong-contrast formalism, as detailed in Section 1 of \href{url}{Supplement 1}.) 
We consider a macroscopically large two-phase composite specimen in three dimensions embedded inside an infinitely large reference phase $q$ \cite{Re08a,torquato_nonlocal_2021}.
For simplicity, we take the phase $q$ to be the matrix phase (i.e., $q\hspace{-2pt}=1,2)$ and assume that phases 1 and 2 are nonmagnetic and dielectrically isotropic with real-valued and frequency-independent dielectric constants. 
These assumptions imply the linear dispersion relation in the reference phase [i.e., $\fn{k_q}{\omega}\equiv\abs{\vect{k}_q(\omega)} = \sqrt{\varepsilon_q}\omega/c$], where $c$ is the speed of light in vacuum, and thus we henceforth do not explicitly indicate the $\omega$ dependence.

The general nonlocal strong-contrast expansion is a series expansion of the linear fractional form of the tensor $\fn{\tens{\varepsilon}_e}{\vect{k}_q}$,  given as
\begin{align}
    &\phi_p \tens{L}_p^{(q)}\cdot 
    \qty(
        \qty{\tens{I} + \tens{D}^{(q)}  \cdot \qty[\fn{\tens{\varepsilon}_e}{\vect{k}_q} - {\varepsilon}_q \tens{I}]} 
        \cdot 
        \qty[\fn{\tens{\varepsilon}_e}{\vect{k}_q} - {\varepsilon}_q \tens{I}]^{-1}
    ) \cdot \phi_p \tens{L}_p^{(q)}
    \nonumber \\
    &=
    \phi_p \tens{L}_p^{(q)} 
    - \sum_{n=2}^\infty \fn{\tens{\mathcal{A}}_n^{(p)}}{\vect{k}_q},
    \label{eq:str-exp-main}
    \end{align}
    where $p(\neq q)$ indicates the polarized phase, $\tens{L}^{(q)}_p$ is the expansion parameter defined as
    \begin{align}
        \tens{L}^{(q)}_p \equiv & \qty({\varepsilon}_p - {\varepsilon}_q) 
        \qty[\tens{I} + \tens{D}^{(q)} \qty({\varepsilon}_p - {\varepsilon}_q)]^{-1},  \label{eq:L-tensor-general-main}
    \end{align}
    and $\fn{\tens{\mathcal{A}}_n^{(p)}}{\vect{k}_q}$ is a wave-vector-dependent second-rank tensor that is a functional involving the set of correlation functions $S_1^{(p)}, S_2^{(p)}, \ldots, S_n^{(p)}$ and products of the principal part of the dyadic Green's function $\fn{\tens{H}^{(q)}}{\vect{r}}$; see Section 1 of \href{url}{Supplement 1}. 
    The series expansion [\eqref{eq:str-exp-main}] has four salient features.
    First, \eqref{eq:str-exp-main} is derived from a spatially nonlocal averaged constitutive relation \cite{agranovich_spatial_2006,chebykin_Nonlocaleffectivemedium_2011, wang_achieving_2018}, resulting in an expansion that is valid from the long- to intermediate-wavelength regimes. 
    Second, \eqref{eq:str-exp-main} exactly treats multiple scattering to all orders at a given incident wave vector $\vect{k}_q$ when the nonlocal homogenization theory is valid because the terms $\fn{\tens{\mathcal{A}}_n^{(p)}}{\vect{k}_q}$ for $n=2,\ldots$ in \eqref{eq:str-exp-main} explicitly account for complete microstructural information (the infinite set of $S_2, S_3, \ldots$) to infinite order \cite{torquato_nonlocal_2021}; see Section 1 of \href{url}{Supplement 1} for details. 
    Third, the choice of the shape of the infinitesimal exclusion region (i.e., $\tens{D}^{(q)}$) leads to a different expansion parameter $\tens{L}_p^{(q)}$ that determines the convergence properties.
    Thus, unlike standard multiple-scattering theories \cite{Sh95, Ts01,vynck_light_2021}, here one can naturally `tune' the general series expansion to obtain distinctly different approximations suited for certain classes of microstructures.
    Fourth, the left side of \eqref{eq:str-exp-main} is a linear fractional transformation of $\fn{\tens{\varepsilon}_e}{\vect{k}_q}$ rather than $\fn{\tens{\varepsilon}_e}{\vect{k}_q}$ itself, which leads to the rapid convergence of the strong-contrast expansions so that its lower-order truncations approximate well higher-order functionals (i.e., multiple scattering) of the exact series to all orders in terms of lower-order diagrams \cite{torquato_nonlocal_2021}.

    \subsection{Multiple-Scattering Approximations for Layered Media}
    \label{sec:theory-layered}
    We can now extract from the exact strong-contrast expansion [\eqref{eq:str-exp-main}] accurate multiple-scattering approximations for layered media by truncating the expansion at the $n$-point level. 
    We focus here on such a formula at the two-point level that depends on the spectral density $\spD{\vect{k}}$ because it is still accurate and easy to compute.
    For simplicity, we restrict ourselves to normally incident waves  (Fig. \ref{fig:schematics}), and thus the effective dielectric constant now depends on the wavenumber $k_q$. 
    We outline the derivation here and provide details in Section 2 of \href{url}{Supplement 1}.
     
    Since layered media have rotational symmetry about the $z$-axis and translational symmetry in the $x\hspace{-2pt}-\hspace{-2pt}y$ plane, the spectral density can be expressed as 
    \begin{align}
        \spD{\vect{k}} = (2\pi)^2 \fn{\delta}{k_x} \fn{\delta}{k_y} \spD{k_z},
    \end{align}
    where $\fn{\delta}{k}$ is the one-dimensional Dirac delta function, and $\spD{k_z}$  is the spectral density of 1D two-phase media.  
    For 1D packings of identical hard rods of radius $a$ and packing fraction $\phi_2$, $\spD{k_z}=\phi_2 [2\sin^2(k_z a)]/({k_z}^2a) \fn{S}{k_z}$ \cite{To02a,To16a}, where $\fn{S}{k_z}$  is the structure factor of the rod centers.
    Due to these symmetries, when applying the series [\eqref{eq:str-exp-main}] to layered media, we utilize the feature 3 discussed in Section 2.\ref{sec:general} by choosing a disk-like exclusion region normal to the $z$-axis \cite{torquato_nonlocal_2021}, leading to 
\begin{align} 
    \tens{D}^{(q)} = & {\varepsilon_q}^{-1} \uvect{z} \uvect{z} , 
    \quad
    \tens{L}^{(q)}_p = \beta_{pq} \Big[ \varepsilon_p   \qty(\tens{I}-\uvect{z}\uvect{z}) 
    + \varepsilon_q  \uvect{z}\uvect{z} \Big], \label{eq:D-strat}
\end{align}
where  $\uvect{z}$ is a unit vector along the $z$-direction and $\beta_{pq}$ is the one-dimensional counterpart of the \emph{dielectric polarizability}, defined as $\beta_{pq} \equiv  1-\varepsilon_q/\varepsilon_p$.
Here, $\tens{L}^{(q)}_p$ is obtained by substituting $\tens{D}^{(q)}$ in \eqref{eq:D-strat} into \eqref{eq:L-tensor-general-main}.
Using \eqref{eq:D-strat} and the assumption of normal incidence, the general expression for $\fn{\tens{\mathcal{A}}_2^{(p)}}{k_q}$ simplifies as 
\begin{align}
    \fn{\tens{\mathcal{A}}_2^{(p)}}{k_q}
    =&
    (\varepsilon_p \beta_{pq} )^2
    \frac{\F{k_q}}{\varepsilon_q}  
    (\tens{I}-\uvect{z}\uvect{z}),
    \label{eq:A2-strat-3} \\
    \F{k}
    = &
    \frac{{k}^2}{\pi} \mathrm{p.v.}\int_{0}^{\infty} \dd{q_z} 
    \frac{\spD{q_z}}{{q_z}^2 - {(2k)}^2}
    +
    \frac{i k}{4} [
    \spD{0} +  \spD{2k}
    ]   \label{eq:F-strat-Fourier},
\end{align}
where $\mathrm{p.v.}$ stands for the Cauchy principal value.
Note that $\F{k_q}$ is the nonlocal attenuation function for 1D two-phase media, whose 2D and 3D counterparts were derived in Ref. \cite{torquato_nonlocal_2021}.

Applying \eqref{eq:D-strat} and \eqref{eq:A2-strat-3}  into the second-order truncation of the series [\eqref{eq:str-exp-main}] yields
\begin{align}
    &
    \frac{(\phi_p \varepsilon_p\beta_{pq})^2}{\varepsilon_q (\fn{\varepsilon_e^\perp}{k_q}/\varepsilon_q - 1)}
    (\tens{I}-\uvect{z}\uvect{z})
    + 
    \frac{(\phi_p \varepsilon_q\beta_{pq})^2}{\varepsilon_q (1- \varepsilon_q / \fn{\varepsilon_e^z}{k_q})}        
    \uvect{z}\uvect{z}
    \nonumber \\
    =&
    \varepsilon_p  \beta_{pq} [\phi_p - (\varepsilon_p \beta_{pq} )
    \F{k_q}/\varepsilon_q ](\tens{I}-\uvect{z}\uvect{z}) + \phi_p \varepsilon_q  \beta_{pq} \uvect{z}\uvect{z}, \label{eq:2pt-strat}
\end{align}
where we have decomposed the effective dielectric constant tensor into two orthogonal components $\fn{\varepsilon_e ^\perp}{k_q}$ and $\fn{\varepsilon_e ^z}{k_q}$ for the transverse and longitudinal polarizations, respectively, as follows:
$\fn{\tens{\varepsilon}_e}{\vect{k}_q} = \fn{\varepsilon_e ^\perp}{k_q} \qty(\tens{I}-\uvect{z}\uvect{z}) + \fn{\varepsilon_e^z}{k_q}\uvect{z}\uvect{z}$.
Now \eqref{eq:2pt-strat} provides two independent approximations:
\begin{align}
    \fn{\varepsilon_e^\perp}{k_q} =& \varepsilon_q \qty[ 
        1 + \frac{{\phi_p}^2 (\varepsilon_p/\varepsilon_q)\beta_{pq}}{\phi_p - (\varepsilon_p \beta_{pq}) \F{k_q} /\varepsilon_q}
    ],
    \label{eq:eps-eff-strat_perp}
    \\
    \fn{\varepsilon_e^z}{k_q} =& \varepsilon_q (1 - \phi_p\beta_{pq})^{-1}.
    \label{eq:eps-eff-strat_z}
\end{align}
Note that $\fn{\varepsilon_e^\perp}{k_q}$ is dependent on the incident wavenumber $k_q$ and is complex-valued if $\spD{0}+\spD{2k_q} > 0$, implying that the media can be lossy due to forward scattering and backscattering from inhomogeneities in the local dielectric constant.
By contrast, $\fn{\varepsilon_e^z}{k_q}$ is independent of $k_q$, reflecting the fact that a traveling longitudinal wave cannot exist under our assumptions.
Hence, we focus on $\fn{\varepsilon_e^\perp}{k_q}$ in the rest of this work.   
In the static limit, Eqs. (\ref{eq:eps-eff-strat_perp}) and (\ref{eq:eps-eff-strat_z}) reduce to  the arithmetic and harmonic means of the dielectric constants, respectively:
\begin{align}
    \fn{\varepsilon_e^\perp}{0} =& \E{\varepsilon}\equiv \phi_p \varepsilon_p + \phi_q \varepsilon_q ,
    &\fn{\varepsilon_e^z}{0} =& (\phi_p/\varepsilon_p + \phi_q /\varepsilon_q)^{-1}.
    \label{eq:eps-eff-strat_perp-static}
\end{align}
Interestingly, these static results are exact for any microstructure  \cite{To02a}.
Renormalization of the reference phase for the optimal convergence (Section 2 of \href{url}{Supplement 1}), equivalent to using the effective Green's function in Ref. \cite{torquato_nonlocal_2021}, yields a \emph{scaled} strong-contrast approximation for disordered layered media:
\begin{align}
    \fn{\varepsilon_e^\perp}{k_q} = \varepsilon_q \qty[ 
        1 + \frac{{\phi_p}^2 (\varepsilon_p/\varepsilon_q) \beta_{pq} }{\phi_p - (\varepsilon_p \beta_{pq} ) \F{ k_q \sqrt{\E{\varepsilon}/\varepsilon_q}}/\E{\varepsilon} }
    ] ,
    \label{eq:scaled-eps-eff-strat}
\end{align}
where $\E{\varepsilon}$ is given in \eqref{eq:eps-eff-strat_perp-static}. 
We henceforth focus on this scaled approximation because it is more accurate than \eqref{eq:eps-eff-strat_perp}, as shown in Fig. S6 in \href{url}{Supplement 1}. 
As shown in Ref. \cite{torquato_nonlocal_2021}, \eqref{eq:scaled-eps-eff-strat} satisfies the Kramers-Kronig relations \cite{Ja90} so that its predictions properly exhibit both \emph{normal dispersion} [i.e., an increase in $\Re[\varepsilon_e^\perp]$ with $k_q$] and \emph{anomalous dispersion} [i.e., a decrease in $\Re[\varepsilon_e^\perp]$ with $k_q$].
Furthermore, satisfying the Kramers-Kronig relations also implies that \eqref{eq:scaled-eps-eff-strat} yields qualitatively accurate predictions, even beyond the validity regime (i.e., $k_q/\rho \lesssim 1.5$); see Sections \ref{sec:results} and 6 on the \href{url}{Supplement 1} for details. 

In the scaled approximation [\eqref{eq:scaled-eps-eff-strat}], the quantity $\F{k_q}$ is generally complex-valued at a given incident wavenumber $k_q$, producing a corresponding $\varepsilon_e^\perp(k_q)$ with a nonnegative imaginary part. 
Following conventional usage, a composite attenuates waves at a given wavenumber if the imaginary part of the effective dielectric constant is positive.
Such attenuation occurs here only because of multiple scattering effects (not absorption).

Importantly, our strong-contrast approximation [\eqref{eq:scaled-eps-eff-strat}] predicts that stealthy hyperuniform layered media can be perfectly transparent (i.e., $\Im[\varepsilon_e^\perp]=0$) in the infinite-volume limit for a finite range of wavenumbers.
Since any stealthy hyperuniform systems completely suppress both forward scattering and backscattering [i.e., $\spD{0}+\spD{2k_1}=0$] for $k_1 < K/2$, $\Im[\F{k_1}]=0$, and substituting this condition into the scaled formula [\eqref{eq:scaled-eps-eff-strat}] yields a perfect transparency interval:
\begin{equation}    \label{eq:trans-regime}
    0 \leq k_1 < K_T \equiv \frac{K}{2\sqrt{\E{\varepsilon}}} = \frac{\rho\pi\chi}{\sqrt{\E{\varepsilon}}},
\end{equation}
where $\E{\varepsilon}$ is the arithmetic mean of the local dielectric constant;  see \eqref{eq:eps-eff-strat_perp-static}.
Perfect transparency implies an infinite localization length within this spectral range, which is consistent with an estimate that the localization length is inversely proportional to what we call the spectral density \cite{izrailev_anomalous_2012}.
We stress that the prediction [\eqref{eq:trans-regime}] is purely theoretical and hence does not rely on simulations or measurements of the spectral density.

\section{Model Microstructures}\label{sec:models}

\begin{figure}[h]
    \centering
    \includegraphics[width=0.45\textwidth]{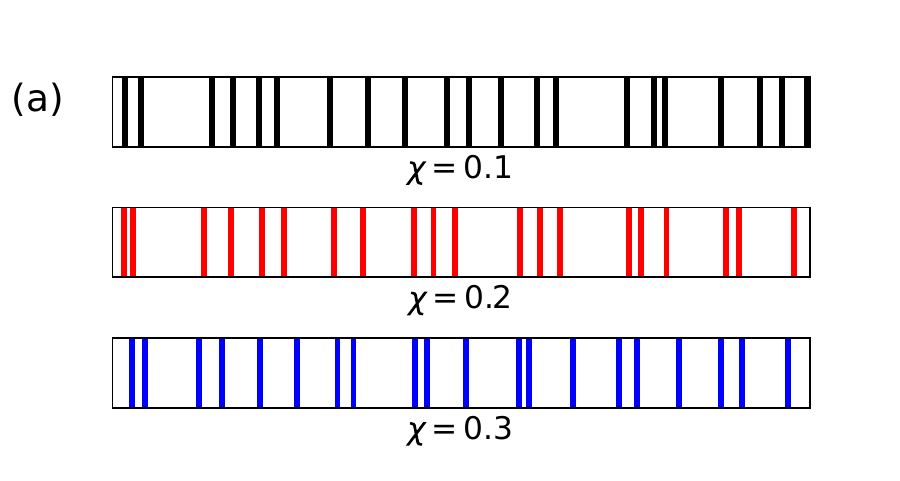}

    \includegraphics[width=0.45\textwidth]{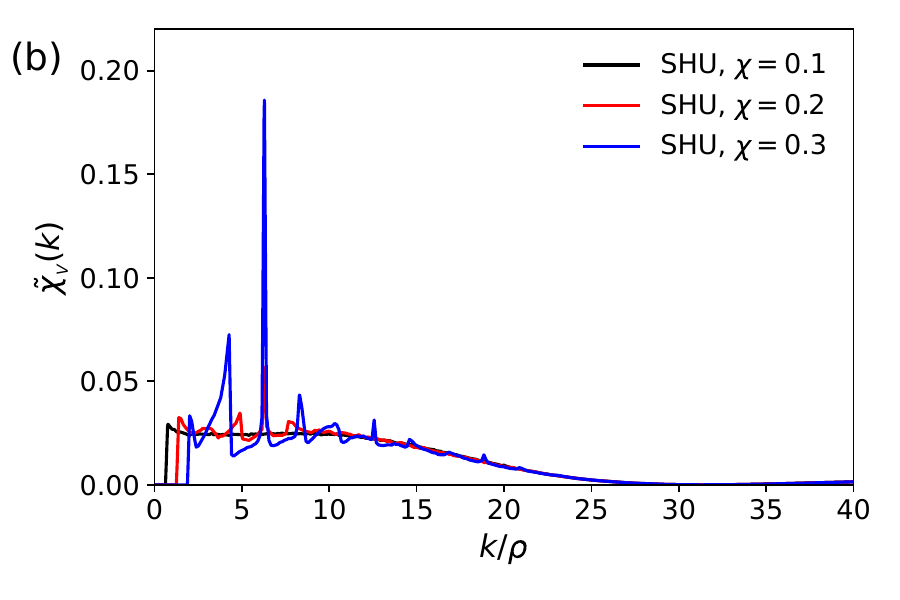}
    \caption{ Disordered stealthy hyperuniform layered media of packing fraction $\phi_2=0.2$ and unit number density $\rho=1$  at three values of $\chi=0.1,0.2,0.3$.
    (a) Representative images at (Top) $\chi=0.1$, (Middle) $\chi=0.2$, and (Bottom) $\chi=0.3$. The particle phases are shown in different colors.
    (b) The spectral densities $\spD{k}$ as functions of dimensionless wavenumber $k/\rho$.  
    \label{fig:models}}
\end{figure}

Stealthy hyperuniform particle systems are defined by a spectral density that vanishes in a finite range of wavenumbers that includes the origin [$\spD{\vect{k}}=0$ for $0<\abs{\vect{k}}\leq K$]. 
The degree of stealthiness $\chi$ is measured by the ratio of the number of the constrained wave vectors in the reciprocal space to the total degrees of freedom, i.e., in one dimension, $\chi = K / (2\pi \rho)$, where $\rho$ is the number density of particles.
For $\chi < 0.5$ in two and three dimensions or $\chi < 1/3$  in one dimension, these stealthy hyperuniform systems are highly degenerate and disordered \cite{Zh15a}.
Thus, we consider 1D  cases in which $\chi$ takes the following values: $\chi=0.1,0.2,0.3$. 
Henceforth, we take the characteristic inhomogeneity length scale $\xi$ to be the mean particle separation $1/\rho$, which is of the order of the mean nearest-neighbor distance $\ell_P$, (i.e., $\ell_P \sim 1/\rho$).
This choice means that the range of validity of our nonlocal theory is $k_1 /\rho \lesssim 1.5$ for the current models, as shown later in Section \ref{sec:results}.  

We numerically generate 1D packings of packing fraction $\phi_2=0.2$ in the following two-step procedure.
First, we generate point configurations of $N$ particles in a periodic  fundamental cell $\mathfrak{F}$ via the \emph{collective-coordinate optimization} technique \cite{Uc04b, Ba08,Zh15a}, which numerically generates ground states with very high-precision of the following potential energy:  
\begin{equation}\label{eq:CC_potential}
\fn{\Phi}{\vect{r}^N} =\frac{1}{V_\mathfrak{F}} \sum_{\abs{\vect{k}}<K} \fn{\tilde{v}}{\vect{k}} \fn{S}{\vect{k}} +  \sum_{i <j} \fn{u}{r_{ij}}, 
\end{equation}
where $\fn{\tilde{v}}{\vect{k}}=1$ for $\abs{\vect{k}}<K$, $V_\mathfrak{F}$ is the volume of $\mathfrak{F}$,  and a soft-core repulsion term \cite{Zh17a} is
\begin{equation} \label{eq:soft}
\fn{u}{r}
=
\begin{cases}
(1-r/\sigma)^2, & r < \sigma,\\
0,&\mathrm{otherwise}.
\end{cases}
\end{equation}
The soft repulsion [\eqref{eq:soft}] is to get the stealthy hyperuniform ground states whose nearest-neighbor distances are larger than the length scale $\sigma$ \cite{Ki20a,  torquato_nonlocal_2021}. 
Finally, we create packings with dielectric constant $\varepsilon_2$  by circumscribing the points by identical rods of radius $a<\sigma/2$ without overlaps \cite{Zh16b}; see Fig. \ref{fig:models}(a).
The parameters used to generate these systems are listed in Section 3 of \href{url}{Supplement 1.}

We compute the spectral density $\spD{k}$  from the generated packings as shown in Fig. \ref{fig:models}(b).
From the long- to intermediate-wavelength regimes ($k/\rho \lesssim 10$),   we clearly see that stealthy hyperuniform packings exhibit a higher degree of correlations as $\chi$ increases.
In the small-wavelength regime ($k/\rho \gg 10$ or, equivalently, $ka \gg 1$),  however, the curves tend to collapse onto a single curve, reflecting the fact that these three systems consist of identical hard rods.

\section{Simulations}
\label{sec:sim}

We validate the accuracy of our predictions on the effective dielectric constant $\fn{\varepsilon_e^\perp }{k_1}$ and the transmittance $T$ by comparing primarily to full-waveform simulations \cite{Taf13} via an open-source FDTD package MEEP \cite{Os10}, although we use the transfer matrix method to compute $T$, as explained in \href{url}{Supplement 1}.
We take the matrix to be the reference phase (i.e., $q=1$) and the particles to be the polarized phase (i.e., $p=2$)  and set the phase contrast ratio as $\varepsilon_2/\varepsilon_1 = 4$.
We measure the transmittance spectra $T$ through the disordered stealthy hyperuniform layered media, which are then compared to the predictions from \eqref{eq:scaled-eps-eff-strat}.
We also directly extract $\fn{\varepsilon_e^\perp }{k_1}$ from the nonlocal constitutive relation $\fn{\varepsilon_e^\perp }{k_1} = \E{\fn{\tilde{D}}{k_e,\omega}}/\E{\fn{\tilde{E}}{k_e,\omega}}$ at a given frequency $\omega$, as was done in Ref. \cite{torquato_nonlocal_2021}; see Fig. S6 of \href{url}{Supplement 1}. 
Here, $\E{\fn{\tilde{D}}{k_e,\omega}}$ and $\E{\fn{\tilde{E}}{k_e,\omega}}$ are the spatial Fourier transforms of the ensemble averages of dielectric displacement field $\E{\fn{D}{x,\omega}}$ and electric field $\E{\fn{E}{x,\omega}}$ at the complex-valued {\it effective wavenumber} $k_e$, respectively; see details in Section 3 of \href{url}{Supplement 1}.

\begin{figure}[t!]
    \centering
    \includegraphics[width=0.5\textwidth]{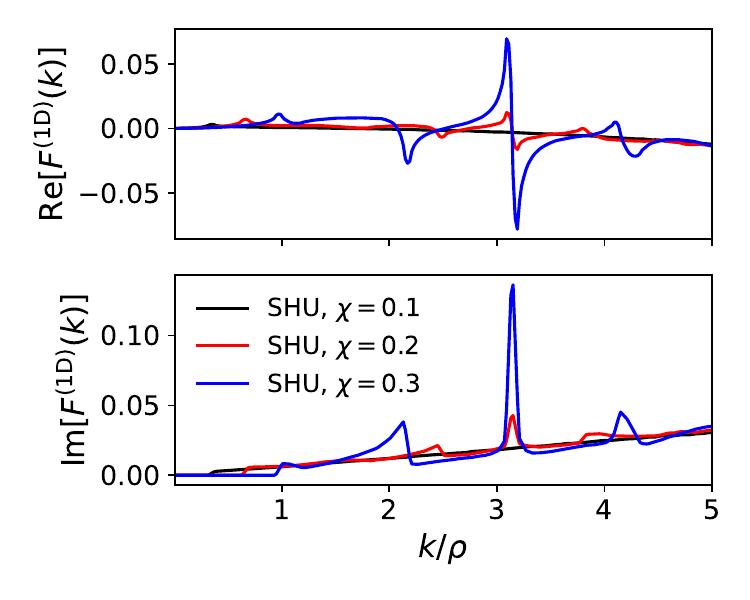}
    \caption{Real (upper) and imaginary (lower) parts of the nonlocal attenuation functions $\F{k}$ as functions of dimensionless wavenumber $k/\rho$ for 1D disordered stealthy hyperuniform layered media of packing fraction $\phi_2=0.2$ at three values of $\chi=0.1,0.2,0.3$. 
    The functions are computed from the spectral densities in Fig. \ref{fig:models} and \eqref{eq:F-strat-Fourier}. \label{fig:F}}
\end{figure}

\section{Results}   \label{sec:results}

\begin{figure}[h]
    \centering
    \includegraphics[width=0.49\textwidth]{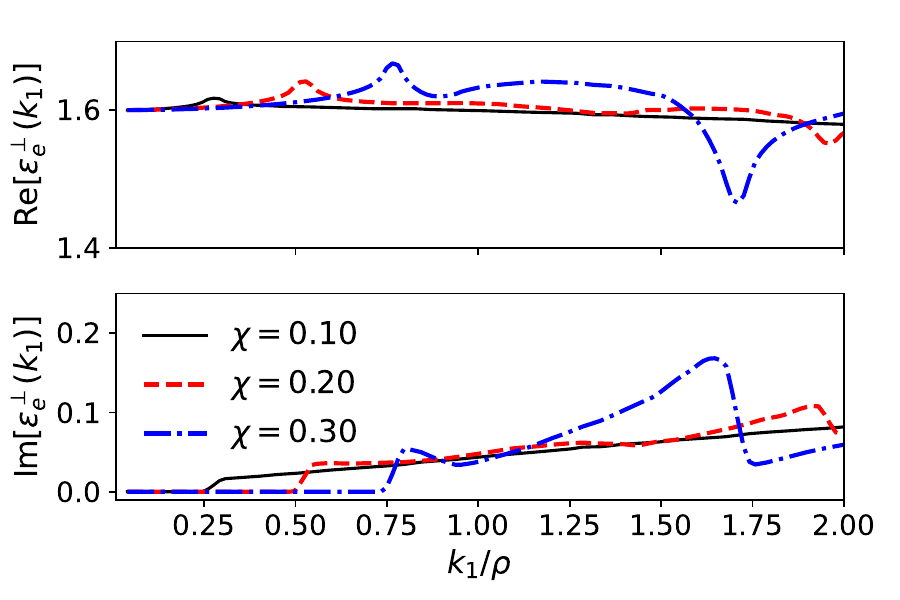}
    \caption{  Predictions of the scaled strong-contrast approximation [\eqref{eq:scaled-eps-eff-strat}] of the effective dynamic dielectric constant $\fn{\varepsilon_e^\perp}{k_1}$ as a function of the dimensionless incident wavenumber $k_1 /\rho$  for disordered stealthy hyperuniform layered media of $\phi_2=0.2$ and $\varepsilon_2/\varepsilon_1=4$ at $\chi=0.1,0.2,0.3$.
    The lower panel is a semilog plot of the imaginary part $\Im[\fn{\varepsilon_e^\perp}{k_1}]$.
    For the effective dielectric constants, our theory is accurate up to $k_1/\rho \lesssim 1.5$; see Fig. S6 of \href{url}{Supplement 1}.}  
    \label{fig:eps}  
\end{figure}

We now show how our multiple-scattering
approximation [\eqref{eq:scaled-eps-eff-strat}] enables us to accurately predict the real and imaginary parts of the effective dielectric constant tensor $\fn{\tens{\varepsilon}_e}{k_1}$ for disordered stealthy hyperuniform layered media for $\chi=0.1,0.2,$ and $0.3$. 
We begin by computing the nonlocal attenuation function $\F{k}$ given in \eqref{eq:F-strat-Fourier} from $\spD{k}$;  see Fig. \ref{fig:F}.
Stealthy hyperuniform layered media can achieve $\Im[\F{k}]=0$ up to a finite wavenumber $k$, leading to the prediction of perfect transparency interval; see \eqref{eq:trans-regime}.

Using the values of $\F{k}$ in Fig. \ref{fig:F}, we then compute the scaled approximation [\eqref{eq:scaled-eps-eff-strat}] for $\fn{\varepsilon_e^\perp}{k_1}$; see Fig. \ref{fig:eps}.
Both real and imaginary parts of these predictions show excellent agreement with the results from FDTD simulations up to $k_1/\rho \lesssim 1.5$  (see Fig. S6 of \href{url}{Supplement 1}).
In Fig. \ref{fig:eps}, the real part of our formula increases with $k_1$ (normal dispersion) within the perfect transparency interval and then decreases with $k_1$ (anomalous dispersion) outside of those intervals.
Such a spectral dependence of $\Re[\varepsilon_e^\perp(k_1)]$ comes from the fact that the strong-contrast formula satisfies the Kramers-Kronig relations \cite{Ja90} (see Section 6 of \href{url}{Supplement 1}), as does the corresponding approximation for 3D statistically isotropic media \cite{torquato_nonlocal_2021}.
Equation (\ref{eq:scaled-eps-eff-strat}) also shows qualitatively accurate dielectric responses even beyond the intermediate-wavelength regime (i.e., $k_1/\rho \gtrsim 1.5$) because the Kramers-Kronig relations closely relate the high-frequency predictions of $\Re[\varepsilon_e^\perp]$ to the accurate predictions of $\Im[\varepsilon_e^\perp]$ in a finite spectral range and vice versa \cite{milton_finite_1997}.

\begin{figure}[h]
    \centering
    \includegraphics[width=0.48\textwidth]{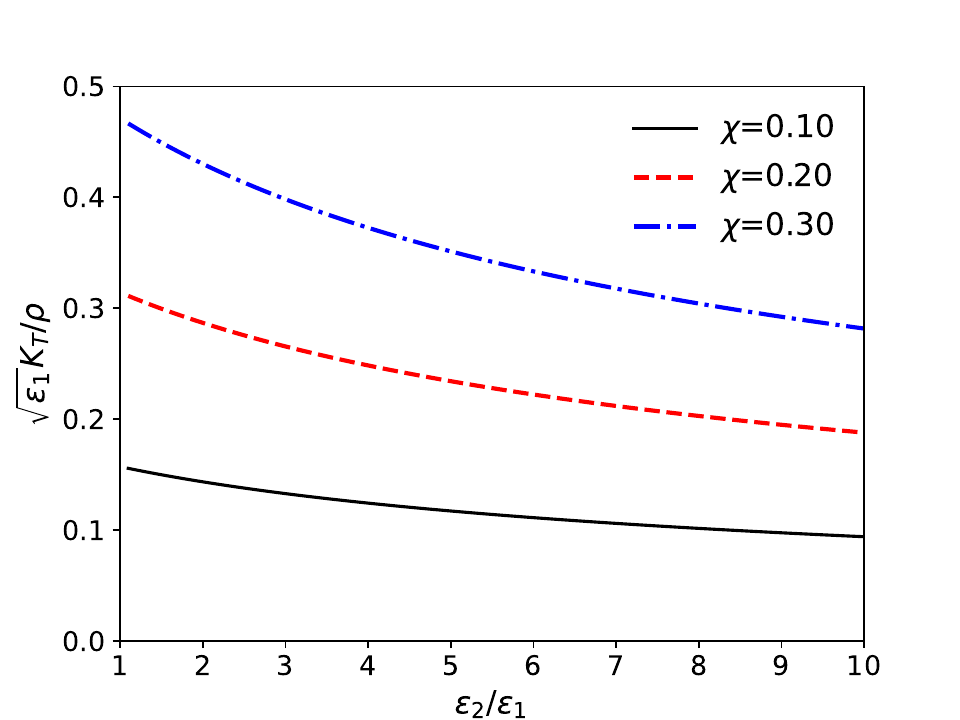}

    \caption{Prediction [\eqref{eq:trans-regime}] of the upper bound $K_T$ of the perfect transparency intervals  of 1D stealthy hyperuniform layered media with $\phi_2=0.2$ and $\rho=1$ as a function of the phase contrast ratio $\varepsilon_2/\varepsilon_1$.
    We consider three $\chi$ values: $\chi=0.1,0.2,$ and $0.3$.
    \label{fig:transparency}
    }
    
\end{figure}

As shown in Fig. \ref{fig:eps}, these composites are perfectly transparent (i.e., $\Im[\varepsilon_e^\perp]=0$) for a wide range of frequencies, as predicted by \eqref{eq:trans-regime}.
At the edges of these transparency intervals, a discontinuous change occurs in $\Im[\varepsilon_e^\perp]$ with $k_1$ because the imaginary part is directly proportional to $\spD{2k_1}$; see $\Im[\F{k}]$ given in \eqref{eq:F-strat-Fourier}.
In higher dimensions, however, such abrupt transitions become increasingly more difficult to achieve, as observed in Ref. \cite{torquato_nonlocal_2021}; see Section 2 of \href{url}{Supplement 1} for detail.
Figure \ref{fig:transparency} depicts how the predicted perfect transparency interval from \eqref{eq:trans-regime} varies with the phase contrast ratio $\varepsilon_2/\varepsilon_1$ for given values of $\chi$, $\phi_2$, and $\rho$.
We numerically demonstrate that these predicted intervals are valid for $1<\varepsilon_2/\varepsilon_1 < 10$; see Fig. S3 of \href{url}{Supplement 1}.

\begin{figure}[h!]
    \centering
    \includegraphics[width=0.45\textwidth]{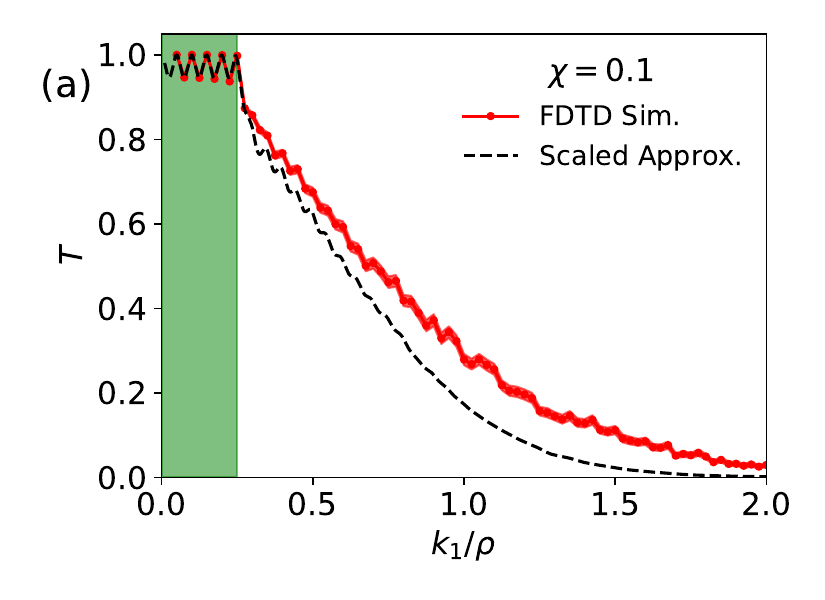}

    \includegraphics[width=0.45\textwidth]{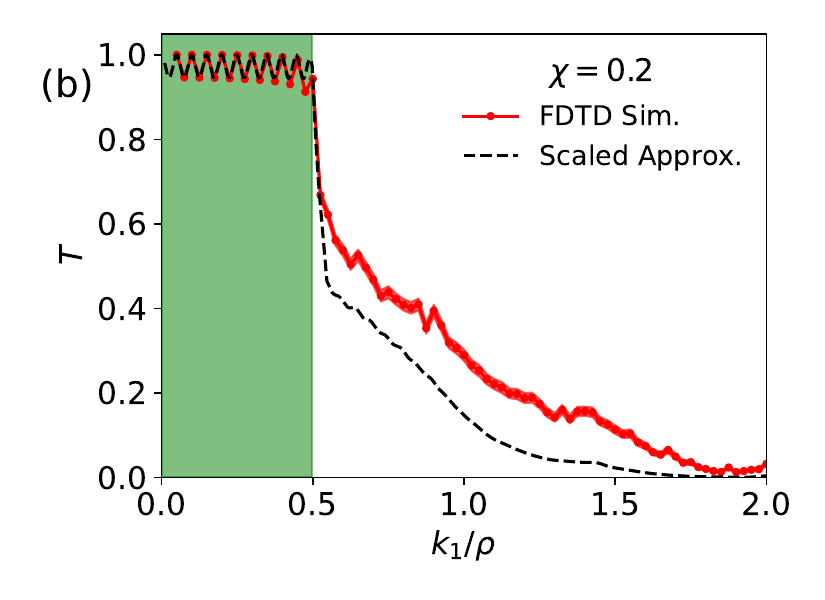}

    \includegraphics[width=0.45\textwidth]{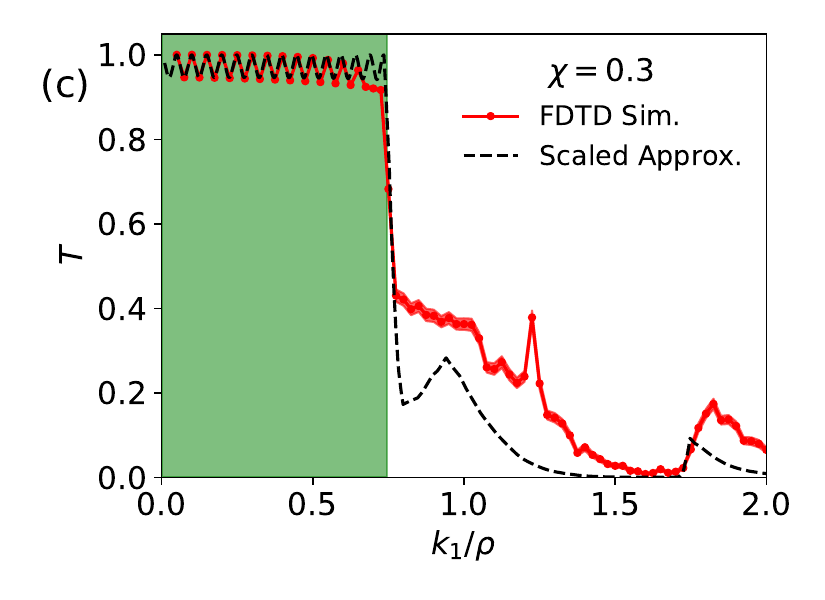}
    \caption{ Transmittance spectra $T$ as a function of the dimensionless wavenumber $k_1/\rho$  for disordered stealthy hyperuniform layered media of packing fraction $\phi_2 \hspace{2pt}(\equiv2\rho a)=0.2$  and phase-contrast ratio $\varepsilon_2/\varepsilon_1=4$ at three values of (a) $\chi=0.1$, (b) $\chi=0.2$, and (c) $\chi=0.3$.
    The predictions are computed from \eqref{eq:T} using the scaled approximation [\eqref{eq:scaled-eps-eff-strat}]. 
    The green-shaded area indicates the predicted transparency intervals [\eqref{eq:trans-regime}]. 
    Within these intervals, our predictions show excellent agreement with simulations, since there is no absorption.
    \label{fig:transmittance}  
    }
\end{figure}

From our scaled approximation [\eqref{eq:scaled-eps-eff-strat}],  we also predict the \emph{normal} transmittance $T$ through a layered medium at $k_1$ by assuming that the system is a homogeneous slab of thickness $L$ with an effective dielectric constant $\fn{\varepsilon_e^\perp}{k_1}$ and is optically thin so that waves inside it can interfere coherently. 
To estimate $T$, we use an Airy formula \cite{yeh_optical_2005-1} of transmittance for a lossy homogeneous slab with absorption:  
\begin{align}   \label{eq:T}
    T &= \abs{\frac{- \sqrt{\varepsilon_e^\perp} ~ t^2 \exp(i \sqrt{\varepsilon_e^\perp} k_1 L)}{1-r^2 \exp(2i \sqrt{\varepsilon_e^\perp} k_1 L)}}^2,
\end{align}
where $r\equiv (1-\sqrt{\varepsilon_e^\perp})/(1+\sqrt{\varepsilon_e^\perp})$ and $t \equiv 2 /(1+\sqrt{\varepsilon_e^\perp})$, and $\varepsilon_e^\perp$ is given by the  approximation [\eqref{eq:scaled-eps-eff-strat}];  see Section 4 of \href{url}{Supplement 1} for results from \eqref{eq:eps-eff-strat_perp}.
The electric field inside the dielectric layered media attenuates solely due to multiple scattering, but its effective behavior is identical to an exponentially damped wave due to absorption in a lossy homogeneous medium (see Fig. S2 of \href{url}{Supplement 1}).
Hence, we expect \eqref{eq:T} to provide a good approximation of $T$.

We now compare our theoretical predictions for transmission to the corresponding results obtained from FDTD simulations; see Fig. \ref{fig:transmittance}.
Remarkably, our theory very accurately predicts the perfect transparency intervals [\eqref{eq:trans-regime}] i.e., no Anderson localization (green regions in Fig. \ref{fig:transmittance}), because they correctly incorporate multiple scattering at finite wavelengths via the spectral density. 
This observation is noteworthy because extended states in 1D disordered systems are much more difficult to achieve than in higher dimensions \cite{Sh95, aegerter_Coherent_2009, wiersma_disordered_2013, yu_Engineered_2021}.
Within these transparency intervals ($k_1 < K_T$), our theory accurately predicts the small-amplitude periodic oscillations in $T$ around unity, which come from coherent interference of the multiply reflected waves due to the finite system thickness $L$ and, thus, reduces to a constant close to unity when $L$ is much larger than the coherence length of light \cite{nichelatti_Complexrefractiveindex_2002,yeh_optical_2005-1}.
Outside of the perfect transparency intervals, where scattering attenuation is strong, $T$ from the Airy formula [\eqref{eq:T}] provides a lower bound on the simulation results for the reason explained in Section 4 of \href{url}{Supplement 1}.
However, we confirm that our theory still accurately predicts, qualitatively, the spectral dependence of $T$ because \eqref{eq:scaled-eps-eff-strat} yields a physically feasible dielectric response due to the Kramers-Kronig relations.
Within these strong-attenuation intervals, $T$ is increasingly suppressed as $L$ increases and becomes virtually zero for sufficiently large $L$ since $\Im[\fn{\varepsilon_e^\perp}{k_1}] > 0$ if $k_1 > K_T$.  
Stealthy hyperuniformity is required for disordered layered media to possess perfect transparency for a finite range of wavenumbers. 
Indeed, we show that such a perfect transparency interval cannot exist for `non-stealthy'  hyperuniform 1D media, even though these correlated disordered systems anomalously suppress large-scale volume-fraction fluctuations; see the example in Section 7 of \href{url}{Supplement 1}.

\section{Conclusions and Discussion}

In summary, we have theoretically and numerically investigated the effective wave properties, including the effective dynamic dielectric constant tensor $\fn{\tens{\varepsilon}_e}{k_q} = \fn{\varepsilon_e^\perp}{k_q} (\tens{I}-\uvect{z}\uvect{z}) + \fn{\varepsilon_e^z}{k_q} \uvect{z}\uvect{z}$  and transmittance $T$,  of 3D statistically anisotropic two-phase layered media made of 1D disordered stealthy hyperuniform packings.
To predict $\fn{\tens{\varepsilon}_e}{k_q}$ of such exotic disordered models, we derived for the first time a multiple-scattering approximation [\eqref{eq:scaled-eps-eff-strat}] for statistically anisotropic media from the strong-contrast-expansion formalism.
Predictions of both the effective dielectric constant $\fn{\varepsilon_e^\perp}{k_q}$ and transmittance $T$ are in {excellent} agreement with corresponding results obtained from the FDTD simulations up to $k_1 /\rho \lesssim 1.5$.
Remarkably, our predictions of $T$ are virtually identical to the simulation results within the perfect transparency ranges.
Our multiple-scattering approximation is the first closed-form formula that provides a simple but accurate relation between the effective wave properties of 3D layered media and their spectral density that applies well beyond the quasistatic regime.
Beyond the valid range $(k_1 /\rho \gtrsim 1.5)$, \eqref{eq:scaled-eps-eff-strat} can still provide qualitatively accurate and physically realistic predictions of dielectric responses, since it satisfies the Kramers-Kronig relations. 

We applied this newly derived formula [\eqref{eq:scaled-eps-eff-strat}] to disordered stealthy hyperuniform layered media at $\chi=0.1,0.2$, and $0.3$.
It is noteworthy that our multiple-scattering formula [\eqref{eq:scaled-eps-eff-strat}] predicts that these disordered systems are perfectly transparent in the infinite-volume limit for a finite ratio $\varepsilon_2/\varepsilon_1$, implying no Anderson localization up to a finite wavenumber proportional
to $\chi$, as given by \eqref{eq:trans-regime}.
This observation is remarkable in that such extended states in 1D  disordered systems are more difficult to achieve than in higher dimensions \cite{Sh95, aegerter_Coherent_2009, izrailev_anomalous_2012, wiersma_disordered_2013, yu_Engineered_2021}.
If the localization length is actually finite, we expect that it will be extremely large for sufficiently small  $\varepsilon_2/\varepsilon_1$ compared to any practically large sample size in the transparency interval in such stratified media, as will be reported elsewhere \cite{Kl_local_2023}.
In contrast, for disordered non-stealthy layered media, hyperuniform or not, our theory shows that there is no spectral range of perfect transparency, implying that localization emerges as the system size grows.

Our findings also have important practical implications. 
For example, we clearly demonstrate that disordered stealthy hyperuniform layered media can be employed as low-pass filters that transmit waves up to a selected wavenumber.
Furthermore, combining our theory with the capabilities to generate media with a prescribed spectral density \cite{Uc04b, Ba08, Zh15a, Ch18a} enables an inverse-design approach \cite{To09a} to engineer and fabricate layered dielectric materials with novel wave properties.   
One possible design is a layered medium satisfying $\spD{k}=0$ for $k<\epsilon$ and $2k_L < k < 2k_U$, which transmits waves within a narrow spectrum $k_L < k_1 < k_U$.
Notably, such computationally designed layered media as well as other 1D disordered models can be readily fabricated via vacuum deposition \cite{affinito_new_1996},  spin-coating \cite{lee_LayerbyLayer_2001}, and 3D printing techniques \cite{tumbleston_continuous_2015}.
Thus, our results offer promising prospects for engineering novel optoelectronic devices.

We note that our formalism can be extended to cases of obliquely incident wavevectors  $\vect{k}_q$, in which both effective dielectric constants $\fn{\varepsilon_e^\perp}{\vect{k}_q}$ and $\fn{\varepsilon_e^z}{\vect{k}_q}$ become important.
Moreover, it also will be interesting to extend our theory to lossy dielectric or metallic phases whose dielectric constants are now frequency-dependent and complex-valued.

\begin{backmatter}
    \bmsection{Funding}
    Army Research Office (W911NF-22-2-0103).  
    
    \bmsection{Acknowledgments}
    The authors thank Michael Klatt for helpful discussions.
    Simulations were performed on computational resources managed and supported by the Princeton Institute for Computational Science and Engineering (PICSciE).

    \bmsection{Disclosures}
    \noindent The authors declare no conflicts of interest.
    
    \bmsection{Data availability} Data underlying the results presented in this paper are not publicly available at this time but may be obtained from the authors upon reasonable request.
    
    \bmsection{Supplemental document} See Supplement 1 for supporting content.
    \end{backmatter}


\end{document}